\documentclass[12pt]{iopart}
\usepackage{graphicx}

\begin{document}

\newcommand{\gguide}{{\it Preparing graphics for IOP Publishing journals}}

\title[Shape reconstruction and height fluctuations of RBCs using defocusing microscopy]{Shape reconstruction and height fluctuations of red blood cells using defocusing microscopy}

\author{L Siman$^1$, P M S Roma$^1$, F T Amaral$^2$, U Agero$^1$ and O N Mesquita$^1$}

\address{$^{1}$Departamento de F\'\i sica, $^{2}$Programa de P\'os
Gradua\c{c}\~ao em Engenharia El\'etrica, Universidade Federal de
Minas Gerais, Caixa Postal 702, CEP 31270-901, Belo Horizonte, MG,
Brazil} \ead{mesquita@fisica.ufmg.br} \vspace{10pt}
\begin{indented}
\item[]October 2014
\end{indented}

\begin{abstract}
In this paper the bright-field defocusing
microscopy (DM) technique is presented. DM is able to obtain quantitative
information of each plane/surface of pure phase objects, as live unlabeled cells, and its
application to red blood cells (RBCs) is demonstrated. Based on
contrast, simple methods to obtain thickness profile and three
dimensional (3D) total reconstruction of RBCs are proposed and the actual height profiles of upper and lower surface-membranes
(lipid bilayer$/$cytoskeleton) of discocyte and stomatocyte red cells
are presented as examples. In addition, using the mean square
contrast fluctuation and modeling the RBC membranes fluctuations
spectra as dependent of a bending modulus $(\kappa_c)$, a surface
tension $(\sigma)$ and a confining potential $(\gamma)$ term, slowly
varying quantities along the cell radius, a genetic algorithm (GA) is used and the
radial height fluctuations of each surface-membrane are accessed,
separately. The radial behaviors of $\kappa_c$, $\sigma$ and
$\gamma$ are also obtained, allowing the discussion of physical
aspects of the RBC membrane.
\end{abstract}

%
%
%

\section{Introduction}

The links between cells mechanics and human diseases have been
subject of considerable scientific research effort for a number of
decades. For example, pathological conditions affecting red blood
cells (RBCs) can lead to alterations to the cell's shape \cite{diez2010}. Also,
changes in RBC membrane properties have influence in the cell
deformability and alters blood circulation \cite{diez2010}. In that
sense, there is an increasing interest in the development of new
techniques able to study mechanical properties of cells.

Optical microscopy has been shown to be a valuable tool to study
biomechanics of living cells, since it can be performed with minimal
perturbation in the cells. The aim of this work is to present new
methods to measure optical and mechanical characteristics of RBCs
using defocusing microscopy
(DM)\cite{bira2003,coelho2005,coelho2007,leo,giuseppe, paula, sebastian_jbo, mariajose}, a simple,
yet powerful optical microscopy technique capable to access
information of each surface-membrane (lipid bilayer/cytoskeleton) of
living cells. We will show, in the case of an adhered red cell, that
information along the $x,y$ position of the upper surface-membrane, free to fluctuate, can be
separately obtained from information along the $x,y$ position of the lower surface-membrane,
which is adhered to the substrate.

Since the contrast of cells without the addition of exogenous
contrast agents is very weakly viewed through a standard
bright-field microscope, the traditional optical microscopy
techniques used to observe living cells are phase-contrast and
Nomarski \cite{zernicke1, nomarski}. These techniques provide
qualitative information or require major calibration to provide the
object's phase map. A number of other optical microscopy
techniques capable to obtain full-field quantitative phase
information have been recently developed \cite{gureyev97,nugent1,
nugent2, popescu2008}. A recent review about these techniques
applied to red blood cells can be found in \cite{kononenko}. As an
example, shape and membrane fluctuations of red blood cells
have been measured using diffraction phase microscopy (QPM)
\cite{popescu2008, pnas2010, popescu2010}, an interferometric
imaging technique that provides quantitative maps of the optical
paths across living cells. DM is a bright-field microscopy technique
also based on interferometric imaging, but with the advantage of
having a simpler experimental set up than QPM, besides being able to
obtain shape and membrane fluctuations of each surface-membrane of
living cells, separately.

Transparent objects that would be invisible in a standard
bright-field optical microscope can turn visible by defocusing the
microscope, which occurs because the act of defocusing introduces a
phase difference between the diffracted and non-diffracted orders.
By recording images at two different focal positions, one can get
information about the phase of the optical electric field and obtain
the object´s full-field phase map. The formalism of Transport
Intensity Equation \cite{teague} has been used for this aim
\cite{gureyev97,nugent1, nugent2}, but in its present form there are no
explicit phase terms considering the distance between the focal
plane and the diffracting surfaces, in a way that the complete
characterization of all interfaces of a phase object is not
feasible. A defocusing technique
has been recently applied for $3$D imaging of cells using a phase
contrast microscope under white light illumination, with transverse
resolution of $350$ nm and axial resolution of $900$ nm
\cite{popescunature}. This technique cannot resolve surfaces
separated by an axial distance smaller than $900$ nm, which is the
case of most RBCs. Strikingly, the defocusing microscopy technique
presented here can resolve surface-membranes of RBCs separated by
axial distances down to $300$ nm, such that cells subject to isotonic, hypertonic and hypotonic solutions can be fully
reconstructed \cite{paula}.

To determine the DM contrast caused by a red cell, a light electric
field is propagated through an infinity corrected defocused
microscope model using the Angular Spectrum propagation formalism
\cite{wolf}. Our approach is limited by the paraxial
approximation, such that the problem is treated within Fresnel
diffraction theory and neglects light polarization. Using the
contrast, simple methodologies to retrieve refractive index, surface
area, volume and three dimension (3D) thickness reconstruction of
red cells are proposed. Also, with the use of a simple
computational method the height profiles of the cell upper and lower
surface-membranes are separately determined \cite{paula}. Here a new
application in RBC is presented, showing the method's potential for
studying the deformation in red cells shape caused by diseases.
Due its simplicity and accessibility, the developed method can be
adopted by non-specialists.

An important measurement when studying cell membranes is the RMS
displacement of the membrane height fluctuations $u_{rms} =
\sqrt{u^2}$, where $u$ is the cell normal displacement around the equilibrium position $h$. In
red cells this phenomena is known as RBC flickering and was first
quantified in $1975$ by Brochard and Lenon \cite{brochard}. More recent models use the well established fact that the RBC surface is composed of an outer
lipid bilayer and an inner cytoskeleton, a quasi-two-dimensional
network of spectrin proteins \cite{safran2007} sparsely
connected to the lipid membrane through specialized proteins, such that this coupled membrane is characterized by effective
elastic moduli: bending modulus $k_c$, tension $\sigma$ and
confining potential $\gamma$, resulting in a
fluctuation spectrum in the planar approximation given by,
\begin{eqnarray}\label{espectro1}
<{|u(\vec{q})|}^2> = \frac{k_{B}T}{\kappa_cq^4 + \sigma q^2 + \gamma},
\end{eqnarray}
where $q$ is the fluctuation wavenumber, $T$ is the absolute
temperature and $k_B$ is the Boltzmann constant. If non-thermal
noise due to detachment of the cytoskeleton from the lipid bilayer
caused by ATP interaction is considered, $T$ is replaced by an
effective temperature, which is higher than the bath temperature
\cite{pnas2010,safran2007,korenstein1991}. The fluctuation spectrum
of equation \ref{espectro1} has been used for analysis of the flickering
phenomena of RBCs \cite{giuseppe,cicuta2009}. In \cite{giuseppe} we have used it to fit the DM mean square contrast fluctuation data measured only in at the center of the cell, showing the technique's ability to access membrane fluctuations, $k_c$ and $\gamma$ of each surface-membrane of RBCs. Here we also assume this spectrum and with the use of a genetic algorithm (GA), the height
fluctuations $u_{rms}$ along the radius of each red cell surface-membrane are
extracted from DM fluctuation contrast data. The radial behaviors of
the elastic moduli are also extracted, allowing the discussion of
mechanical aspects of RBC surface-membranes. Although the choice of
a fluctuation spectrum is a requisite, the ability to obtain
$u_{rms}$ do not dependent on the chosen model. By using the planar
membrane model of equation \ref{espectro1} a more direct comparison with
results obtained by other techniques can be carried out. As an
example, the area compressibility moduli $KA$ obtained here from our
$\gamma$ results are in agreement with the values measured for whole
RBCs using DPM technique \cite{popescu2010}. Finally, in order to
test the validity and accuracy of DM technique, a set of control
experiments using phase gratings were performed and are presented in
the Supplemental material. The grating's height profiles and
refractive indexes were measured using well establish methods and
compared with the results obtained using DM technique.

\subsection{Defocusing microscopy}

\begin{figure}[t!]
\centering{\includegraphics[width=16pc]{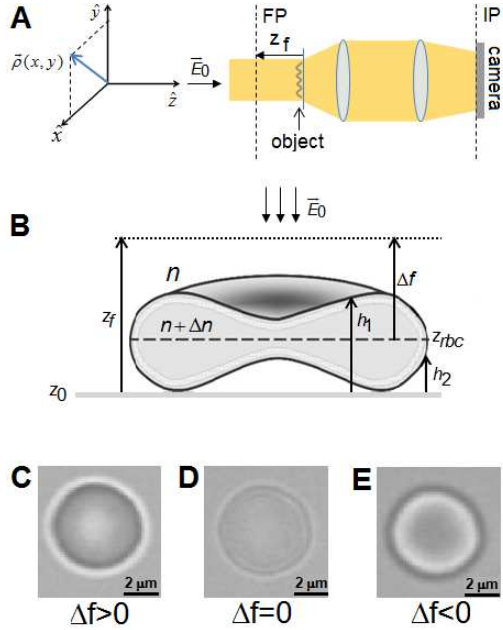}}
\caption{(A) Scheme of a bright-field microscope with infinity
corrected optics. If the optical axis is defined as the z axis, with
its origin at the coverslip position, the defocus distance $z_f$ is
given by the distance between the object and the objective focal
plane (FP).(B) Model for a RBC observed through a defocused microscope.
The cell has a refractive index of $n + \Delta n$, where $n$ is the
immersion medium refractive index and $h_1$ and $h_2$ are the distances
between the coverslip and upper and lower membrane, respectively. The distance $\Delta f$ between the cell mean plane ($z_{rbc}$) and the objective focal plane is used during the experiments to quantify the defocus amount.
Three contrast images of a cell in: (C)$\Delta f>0$,
(D)$\Delta f=0$ and (E)$\Delta f<0$.}\label{figure1a_1e}
\end{figure}

A bright-field microscope with infinity corrected optics is
illustrated in figure \ref{figure1a_1e}(A). Light emitted from a
halogen lamp passes through the microscope condenser system aligned
in K\"{o}hler configuration \cite{wolf}, such that plane waves with
uniform intensity (colour filtered leaving only the appropriate frequency transmitted) lighten the object. After passing through the
specimen both transmitted and diffracted rays are collected by the
objective lens, which together with the tube lens create an
amplified image of the object at the image plane (IP). Assuming the
reference system in figure \ref{figure1a_1e}(A), where $z$ is the
optical axis, the origin can be settle anywhere along it. The origin
is set at the coverslip plane and the defocus distance ($z_f$) is
defined as the distance between the coverslip and the objective
focal plane (FP).
If the observed specimen is a thin pure phase object placed at FP, no
contrast can be seen at IP. When the object is slightly shifted from
FP a contrast $C$ is observed. This contrast is named here as DM
contrast and defined as $C(\vec{\rho})=\frac{I(\vec{\rho}) -
I_0}{I_0}$, where $I(\vec{\rho})$ is the light intensity at a point
$\vec{\rho}$ of the image plane and $I_0$ is the background
intensity of the same plane. It is important to mention that DM
technique does not require the use of an infinity corrected optical
system, but it is fundamental to have a homogenous illumination.

Red cells can be treated as pure phase objects when illuminated by
red light because in this case light absorption is negligible. Therefore the incident light is filtered, letting a transmitted light with $\lambda= (0.660\pm0.010)\;\mu$m. As
shown in figure \ref{figure1a_1e}(B), RBCs can be modeled as objects
composed of an upper and a lower surface-membrane (lipid bilayer $+$
cytoskeleton), separated by a typical distance of $2\;\mu$m at the
rim. Taking the coverslip plane as the $z$ axis origin, $h_1$ is
defined as the distance between the coverslip and the cell upper
membrane, $h_2$ as the distance between the coverslip and the lower
membrane and $H = h_1 - h_2$ as the cell thickness. The cell is
considered to have an uniform refractive index $n_{rbc}=n + \Delta
n$, where $n$ is the immersion medium refractive index. Performing
the propagation for an electric field $E_0$ with wave number
$\vec{k} = k_0\hat{z}$ (Supplementary material), for first order
diffraction, the contrast $C(\vec{\rho})$ is given by,
\begin{eqnarray}\label{crbc0}
C(\vec{\rho})=\frac{2\Delta nk_{0}}{\sqrt{S}}\Bigg\{\sum_{\vec{q}}
\Bigg[h_2(\vec{q})\sin\Bigg(\frac{(z_f-h_2(\vec{\rho}))q^2}{2k}\Bigg) \nonumber\\
-h_1(\vec{q})\sin\Bigg(\frac{(z_f-h_1(\vec{\rho}))q^2}{2k}\Bigg)\Bigg]\sin(\vec{q}\cdot\vec{\rho})\Bigg\}.
\end{eqnarray}
Here $k_0$ is the vacuum illumination light wavenumber, $\Delta n$
is the refractive index difference between the immersion
medium and the RBC, $S$ is the cell surface area and $k=n_{ob}k_0$,
where $n_{ob}$ is the objective immersion medium refractive index.
The terms $h_{1/2}(\vec{q})$ are the spatial Fourier components of
the phase object height profiles
$h_{1/2}(\vec{\rho})=\frac{1}{\sqrt{S}}\sum_{\vec{q}}h_{1/2}(\vec{q})\sin(\vec{q}\cdot\vec{\rho})$.
Considering $\frac{(z_f - h_1)q^2}{2n_{ob}k_0}\ll1$, $\frac{(z_f -
h_2)q^2}{2n_{ob}k_0}\ll1$, and
$\frac{1}{\sqrt{S}}\sum_{\vec{q}}\bigg(h(\vec{q})\sin(\vec{q}\cdot\vec{\rho})q^2\bigg)=
- \bigtriangledown ^{2}h(\vec{\rho})$, we obtain the DM contrast
for a RBC,
\begin{eqnarray}\label{crbc1}
C(\vec{\rho})\simeq\frac{\Delta n}{n_{ob}}\Bigg[\bigg(z_f -
h_1(\vec{\rho})\bigg) {{\bigtriangledown^{2}}h_{1}(\vec{\rho})} -
\bigg(z_f
-h_2(\vec{\rho})\bigg){{\bigtriangledown^{2}}h_{2}(\vec{\rho})}\Bigg].
\end{eqnarray}
The validity of equation \ref{crbc1} can be determined for the interval
that C varies linearly with $z_f$. This approximation will be used
for $3D$ total reconstruction of RBCs since the average shape
involves low wavenumbers $q$. For fluctuation analysis, the complete equation \ref{crbc0} has to be used, as shown below. In DM formulation there is no doubt
over which wavenumber of light to be used, since the phase shift
introduced by each optical element causing the image defocusing is
known. If defocusing is generated by displacing an oil immersion
objective, the phase shift introduced is related to optical path
difference in that medium and the wavenumber to be used in the
defocusing term is $k=k_0n_{ob}=k_0\times 1.51$. Differently, if
defocusing is generated by displacing a dry objective, then
$k=k_0n_{0}=k_0$.

An important characteristic of cell membranes is their height
fluctuation in relation to their average position. Those height
fluctuations induce contrast fluctuations and the mean
square contrast fluctuation ($<(\Delta C)^2>$) contains information
about them, such that the elastics moduli can be
extracted. In this case, each height profile is described by a
constant term $h_{1/2}$ and a time-dependent  fluctuating term $u_{1/2}$, so that
$h_{1/2}(\vec{\rho},t) = h_{1/2}(\vec{\rho}) +
u_{1/2}(\vec{\rho},t)$. The contrast fluctuation $\Delta
C(\vec{\rho},t) = C(\vec{\rho},t) - <C(\vec{\rho},t)>$, where
$<C(\vec{\rho},t)>$ is the mean contrast over time gives,
\begin{eqnarray}\label{deltacrbc1}
<\Delta C^2(\vec{\rho})> = \frac{(2\Delta
nk_{0})^2}{2S}\sum_{\vec{q}}\Bigg[<|{u_{1}(\vec{q})}|^2>
\times\sin^2\Bigg(\frac{z_f -h_1(\vec{\rho})}{2k}q^2\Bigg)\nonumber\\
+<|{u_{2}(\vec{q})}|^2>\sin^2\Bigg(\frac{z_f-h_2(\vec{\rho})}{2k}q^2\Bigg)\Bigg],
\end{eqnarray}
where $<|u_{1/2}^2(\vec{q})|^2>$ are the membranes fluctuation
spectra. Equation \ref{deltacrbc1} has two minima: when $z_f=h_1(\vec{\rho})$ and when $z_f=h_2(\vec{\rho})$. By scanning $z_f$ and measuring $<(\Delta C)^2>$ at the center of RBCs, Glionna et al. \cite{giuseppe} showed that these two minima for $<(\Delta C)^2>$ are displayed and it was possible to determine the elastics constants at the center of the cell, for each membrane, separately.

\section{Materials and methods}

\subsection{Microscopy}

All experiments were conducted on inverted microscopes operating in
bright-field (Nikon TE$300$ and Nikon Eclipse Ti, Nikon Instruments
Inc.,Melville, NY) with a $100X$ oil immersion objective (Nikon Plan
APO DIC H, $1.3$ NA, Nikon). Images were either captured with a CMOS camera
of $640\times480$ pixels, (Silicon Video SV$642$M, EPIX Inc, Buffalo
Grove, IL) at a typical frame rate of $333$fps, or with a CCD camera
of $1390\times1037$ pixels (UP$1800$CL$-12$B, UNIQ Vision Inc, Santa
Clara, CA) at $15$fps, depending on the experiment. The defocusing
distance was controlled by a piezoelectric nanoposition stage
(P$563-3$CD, Physik Instrumente (PI) GmbH Co, Karlsruhe) with
precision of $5$ nm or by using the Nikon Perfect Focus System. The
cells were illuminated with red filtered light
($\lambda=(0.660\pm0.010)\;\mu$m).

\subsection{Sample preparation}

Red blood cell samples were prepared by a standard procedure to
yield discocyte RBCs \cite{leo}, with informed consent from the human
subjects. For each experiment $500\;\mu$L of RBC suspension was
transferred into a microscope coverslip chamber and covered to avoid
water evaporation. All experiments were performed at room
temperature ($T=299$ K).

\subsection{Cell imaging}

For RBCs experiments it is convenient to settle the z axis origin at
the cell mean plane $(z_{rbc})$, so that $z_f = \Delta f +
z_{rbc}$, where $\Delta f$ is the distance between the cell mean
plane and the objective focal plane (figure \ref{figure1a_1e} (B)). Also, $h_1$ and $h_2$ are
defined as the distances between $z_{rbc}$ and the upper and lower
membrane, respectively, and the cell thickness is now given by
$H=h_1+ |h_2|$. The mean plane is easily determined during cell
imaging as the minimal contrast plane along the $z$ axis. In figure
\ref{figure1a_1e}(C,D,E), RBC images for $\Delta f > 0$, $\Delta f =
0$ and $\Delta f< 0$ are presented. For shape results, $42$ RBCs were
imaged for $20$ seconds at defocus positions $\Delta f=2\;\mu$m
($C_1$) and $\Delta f=0$ ($C_2$), at a frame rate of $15$ fps with
the UNIQ camera. For membrane fluctuations results the same $42$ cells were
imaged for $30$ seconds at defocus position $\Delta f= 2 \mu$m and
using a frame rate of $333$ fps with the CMOS camera. It is
important to mention that several cells can be imaged
simultaneously, regarding that they are not in contact with each
other. Using plugins developed by our group, $<C(\vec{\rho})>$ and $<(\Delta
C(\vec{\rho}))^2>$, where the averages are time averages, were obtained for each stack of images. The camera noise was subtracted from all $<(\Delta
C(\vec{\rho}))^2>$ data.

\section{Results and discussion}

\subsection{Thickness profile}

\begin{figure}[t!]
\centering{\includegraphics[width=30pc]{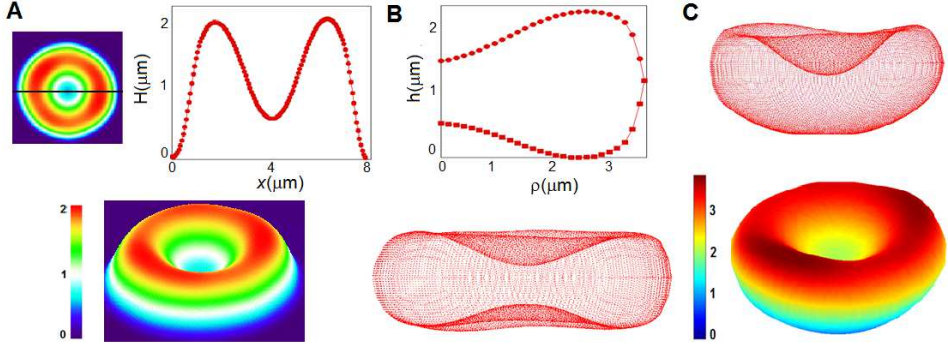}}
\caption{(A)Thickness map of a red cell and
its $3D$ thickness reconstruction. The plot shows the cell
thickness $H$ along the horizontal line drawn in the thickness map.(B)
Radial height profiles for upper (points) and lower (squares) membranes
of a RBC and its $3$D total image.(C)$3$D total image of a stomatocyte RBC in two different views and representation. Color bars in $\mu$m. }\label{figure2a_2c}
\end{figure}

To obtain RBC thickness reconstruction and volume we use equation \ref{crbc1} and subtract
the cell average contrast taken at two different defocus distances,
$<C_1(\vec{\rho})>$ and $<C_2(\vec{\rho})>$. Performing a Fourier transform
(FFT) on it, the laplacian term becomes $-q^2(h_{1}-h_2)$, where
$H(\vec{\rho})= h_{1}-h_2$ is the cell thickness. Dividing it by
$(-q^2)$ and performing an inverse Fourier Transform,
\begin{eqnarray}\label{perfilaltura2}
H(\vec{\rho})=\frac{n_{ob}}{\Delta
n(z_{f_2}-z_{f_1})}
{\mathcal{F}}^{-1}\Bigg\{\frac{{\mathcal{F}}\{<C_1(\vec{\rho}>)-<C_2(\vec{\rho})>\}}{-q^{2}}
\Bigg\},\nonumber\\
\end{eqnarray}
the cell thickness is then obtained. The spatial average intensity of the
entire image $<I>_{\vec{\rho}} =\frac{1}{A_0}\int{I\:d{\vec{\rho}}}$ of a pure phase object is a constant, given by the
background intensity $<I>_{\vec{\rho}} =I_0$. That implies that the
average contrast $<<C>>_{\vec{\rho}}=(<I>-I_0)/I_0=0$ for the entire image. Since the Fourier
transform of the contrast for $q=0$ is $<C>=0$, we disregard this
point in equation \ref{perfilaltura2}, eliminating the problem of the
division by $q=0$. This algorithm (equation \ref{perfilaltura2}) was
previously proposed in \cite{gureyev97} to retrieve the phase map of
pure phase objects. However, since the phase was not explicitly
solved, the object's thickness profile was not obtained. The above
procedure can be easily performed using Image J (Rasband WS. ImageJ, U.S.
National Institutes of Health, Bethesda, Maryland, USA,
imagej.nih.gov/ij/, 1997–2012) \cite{image}. Since the values for
$z_f$, $n_{ob}$ and $\Delta n$ are known, the resultant image is a
thickness map of the observed cell, which corresponds to the cell's
phase map when divided by $\Delta n$. The cell refractive index can
also be determined using DM measurements and the procedure is exposed in \cite{paula}. Here the value of $\Delta n= 0.058\pm0.003$ is
used. Cell volume can be determined as $V= A_{pixel}\times
H$, using the pixel area $A_{pixel}$. A $3D$ thickness image can
be visualized using the $3D$ Surface Plot plug in from Image J
\cite{image}. In figure \ref{figure2a_2c}(A), a red cell thickness map
is shown together with its $3D$ thickness reconstruction. On the right side,
the thickness profile plot along a horizontal line of the thickness map is
presented. The mean volume for the $42$ analyzed cells was
$\overline{V} = 93 \pm 19\; \mu$m$^3$, which is within the range reported by other techniques \cite{rappaz2008}.

\subsection{3D total imaging}

An unique capability of DM is to image the upper and lower surfaces of a phase object, what we call $3D$ total imaging. To determine the height profiles of the cell upper and lower
surface-membranes, separately, the cell thickness profile and the
contrast image for one single defocus position, $C_1(\vec{\rho})$, is
used. Defining the asymmetry between the cell's membranes as
$\delta(\vec{\rho})= h_1 - |h_2|$, for $z_{f}=0$, equation (\ref{crbc1})
can be rewritten as,
\begin{eqnarray}\label{3dtotal_1}
\nabla^2\delta+\frac{\nabla^2H}{H}\delta=-\frac{2n_{ob}}{H\Delta
n}C(0,\vec{\rho}).
\end{eqnarray}
Equation \ref{3dtotal_1} is a non-homogeneous Helmholtz equation with
variable coefficients that is numerically solved for
$\delta(\vec{\rho})$ with the initial condition
$\delta(\vec{\rho})=0$ \cite{paula}. From the returned asymmetry
$\delta(\vec{\rho})$, the surface-membranes height profiles for each
pixel of the contrast image are recovered,
\begin{eqnarray}\label{3dtotal_2}
h_1(\vec{\rho})=\frac{H(\vec{\rho})+\delta(\vec{\rho})}{2}\;\;\;\;\;\;\;\;h_2(\vec{\rho})=\frac{-H(\vec{\rho})+\delta(\vec{\rho})}{2}\nonumber\\
\end{eqnarray}
and the total $3$D imaging is obtained. Using the developed method
the height profiles $h_1$ and $h_2$ for the $42$ analyzed cells
were performed. In figure \ref{figure2a_2c}(B) an example of the radial height profiles $h_1$
(red points) and $h_2$ (red squares) for a RBC, together with its $3$D total
image is presented. In figure \ref{figure2a_2c}(C) different representations of two
views of the $3$D total image of a stomatocyte red cell are
also presented. As one can see, the $3$D total imaging method access the
actual bowl-like shape of the stomatocyte cell, showing that DM technique could be useful
to investigate the relation between red cells deformation
and pathologies, as well as adhesion to substrates. Further details of the method can
be seen in \cite{paula}.

\subsubsection{DM axial resolution}

DM axial resolution is based on the sensitivity of image contrast
measurement, optical contrast of the phase object and mean
curvature of the surfaces visualized. For an estimate, let us
determine the difference in contrast (equation \ref{crbc1}) between the
two surfaces at ${\vec{\rho}=0}$, where the minimum distance between
the membrane-surfaces occurs in normal RBCs,
\begin{eqnarray}\label{ax1}
\Delta C=\frac{\Delta n}{n_{ob}}(h_1 - h_2)(\nabla^2h_1 -
\nabla^2h_2)=\frac{\Delta n}{n_{ob}}H(0)\nabla^2H(0).
\end{eqnarray}
Here $H(0)=h_1(0)-h_2(0)$ is the axial distance between the two
surface-membranes in the center of the cell. The minimum axial
distance that can be resolved using DM is then,
\begin{eqnarray}\label{ax1}
H(0)_{min}=\frac{n_{ob}\Delta C_{min}}{\Delta n\nabla^2H(0)}.
\end{eqnarray}
Considering a contrast sensitivity of $\Delta C_{min}=10^{-2}$, the
oil immersion refractive index $n_{ob}=1.51$ and the typical values
for isotonic RBCs, $\Delta n=0.058$, and
$\nabla^2H(0)=1.7\;\mu$m$^{-1}=1.7\times10^{-3}$ nm$^{-1}$, then
$H_{min}(0)\sim 150$ nm, allowing clear visualization of the two RBC surfaces.

\subsection{Surface-membranes height fluctuation}

In order to access RBC surface fluctuations one can measure the mean
square contrast fluctuation $<(\Delta C)^2>$ at one defocus
distance. Using the continuum version of equation \ref{deltacrbc1},
\begin{eqnarray}\label{deltacrbc_c}
<\Delta C^2(\vec{\rho})>=\frac{(\Delta
nk_{0})^2}{\pi}\int_{q_{min}}^{q_{max}}
qdq\Bigg[<|{u_{1}(q)}|^2>\times\sin^2\Bigg(\frac{(z_f-h_1)q^2}{2k}\Bigg) +\nonumber\\
<|{u_{2}(q)}|^2>\times\sin^2\Bigg(\frac{(z_f-h_2)q^2}{2k}\Bigg)\Bigg],
\end{eqnarray}
and the profiles for $h_1(\vec{\rho})$ and $h_2(\vec{\rho})$ we previously obtained, then the fluctuation power
spectra $|u_{1/2}(\vec{q})^2|$ as in equation \ref{espectro1} and the RMS
height fluctuation of each cell membrane can be determined by
computational methods. A genetic algorithm  was implemented and employed (Supplementary
material) and the results for $u_{1/2(rms)}$, $k_{c1/2}$,
$\gamma_{1/2}$ and $\sigma_{1/2}$ for an average of $42$ RBCs are
presented. For the GA fittings the continuum form of equation
\ref{deltacrbc1} is used. The integration approach is appropriate to
the red cell membrane deformation and the results are not very
sensitive to the exact values of the limits $q_{min}$ and $q_{max}$.
Assuming the elastic constants as slowly varying functions of
$\vec{\rho}$, an adiabatic approximation can be done by assuming a
piecewise constant spectrum. The integration intervals used are
$q=[0.6,12.0]\mu$m$^{-1}$ for the upper membrane and
$q=[1.2,12.0]\mu$m$^{-1}$ for the lower one. The minimum value of
$q$ corresponds to the first zero of the Bessel function for the
cell radius $R=4\;\mu$m (upper membrane) and for a radius of
$R=2\;\mu$m (lower membrane). This difference occurs because the
lower membrane is attached to the substrate for $R>2\;\mu$m (figure \ref{figure2a_2c}(B)),
reducing some possible wavenumbers. The maximum value for $q$ is
determined by the objective numerical aperture, $q\sim12\;rad/\mu$m.

\begin{figure}[t!]
\centering{\includegraphics[width=32pc]{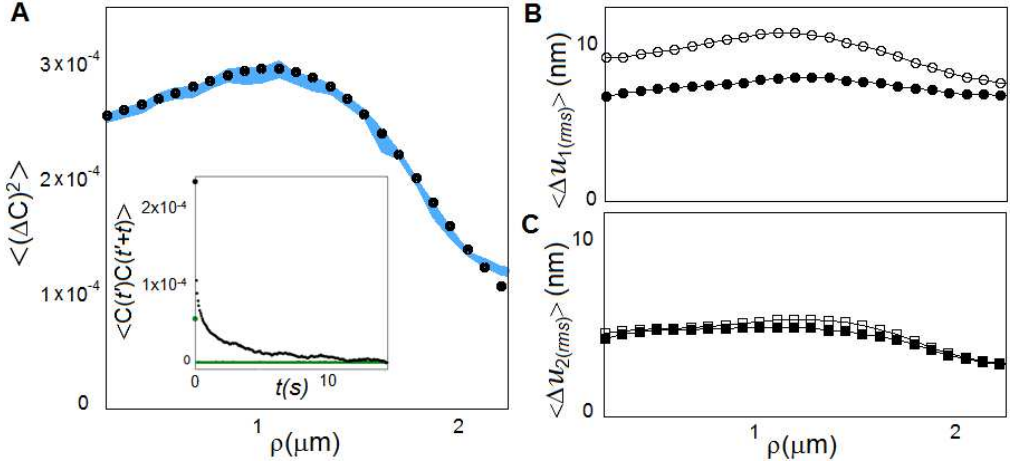}}
\caption{(A) Average of the mean square contrast fluctuation $<(\Delta C)^2>$ at $\Delta f=2\;\mu$m (dots) for $42$ RBCs. The blue area represents the ten final fittings returned by the AG. Inset: time autocorrelation function of contrast on the RBC and on the background, which presents only a shot-noise at $t=0$. (B) RMS height fluctuation of the upper membrane $u_{1(rms)}$ before (empty dots) and after (full dots) correction for cell lateral movement. (c) RMS height fluctuation of the upper membrane $u_{2(rms)}$ before (empty squares) and after (full squares) correction.}\label{figure3a_3c}
\end{figure}

In figure \ref{figure3a_3c} (A) the average for $42$ cells of the mean
square contrast fluctuation $<(\Delta C)^2>$ at $\Delta f=2\;\mu$m, after camera noise subtraction, is displayed. The blue area represents the ten final fittings returned by the AG and the small deviation indicates that the chosen plane membrane model yields a good fitting of the experimental data. On the plot inset, the temporal contrast autocorrelation function along a line in the cell center (black dots) and along a line in the image background (green dots) are shown. As it can be seen, for the sampling time used the background time correlation function consists of a term in $t=0$ (shot-noise) and decays immediately to zero. This shot noise is subtracted from the contrast fluctuations data measured in the cells. We observe that a good signal to noise ratio is obtained down to $<(\Delta C)^2>\sim2\times10^{-6}$, returning a sensitivity of about $1.4$ nm in the measurements of height fluctuations.

The RMS displacement of membrane height fluctuations $u_{1(rms)}$ and
$u_{2(rms)}$ are shown in figure \ref{figure3a_3c}(B-C). Four sets of data
are presented and the ones depicted by empty dots and squares represent
the upper and lower membrane $u_{rms}$, respectively. The membranes
have a similar height fluctuation profile, but the upper membrane
fluctuates more than the lower one, since the lower membrane is
closer to the substrate and adhered to it in a certain region,
confining and restricting membrane fluctuations. The two other
curves (full dots and squares) in the graph represent $u_{1(rms)}$ and $u_{2(rms)}$ after
being corrected for the cell lateral movement. The correction term is
$\sqrt{4-3\cos^2(\theta)}$ and it is fully deduced in the
Supplementary material. In \cite{rappaz2008}, \emph{Rappaz et al.}
suggested a different correction, where the height fluctuation
amplitudes are modified by a factor $\cos(\theta)$, with $\theta$
the angle that the normal to the cell surface makes with the $z$ axis.
Since DM technique can access the height profile of each cell
membrane, our proposed correction might be more accurate.

\subsection{Confining potential $\gamma$}

\begin{figure}[t!]
\centering{\includegraphics[width=25pc]{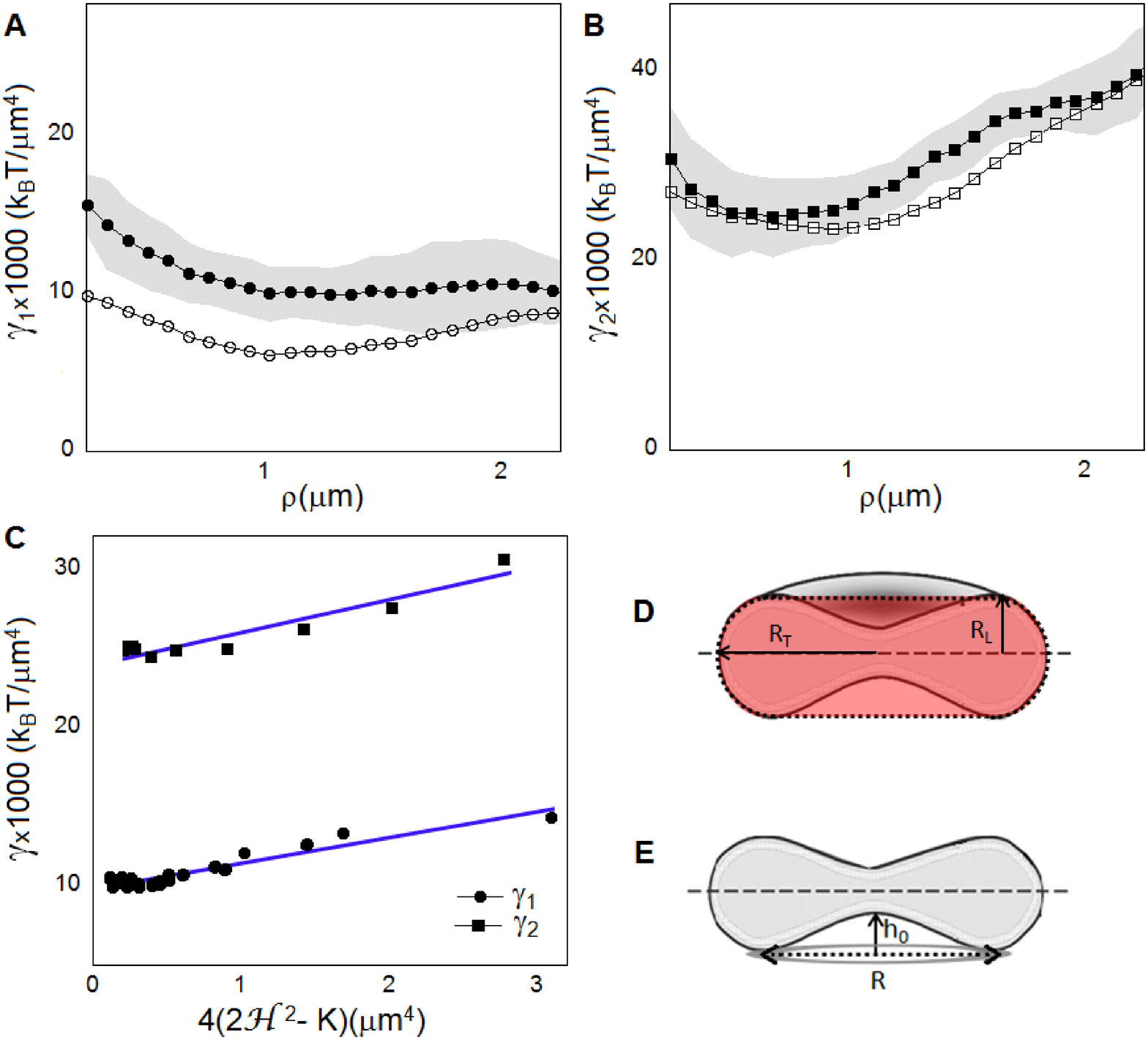}}
\caption{Confining potential for each cell membrane: (A) $\gamma_1$ and (B) $\gamma_2$. The unfilled
symbols represent the non-corrected data and the filled symbols
represent the corrected ones. The grey area is the standard
deviation of the mean value obtained with the ten final runs of the AG. (C) $\gamma$ \emph{versus} $4(2\mathcal{H}_2 - K)$ for each membrane with the
linear fit $\gamma= KA\times 4(2\mathcal{H}^2 -
K) + \gamma_{0}$. The returned values for the
compressibility moduli are $KA_1=(1.6\pm0.1)\times 10^3\;k_BT/\mu
m^2\sim7\;\mu N/m$ and $KA_2=(2.0\pm0.3)\times 10^3\;k_BT/\mu
m^2\sim8\;\mu N/m$.
Also, $\gamma_{0_1}=(9.9\pm0.1)\times 10^3\;k_BT/\mu m^4$ and
$\gamma_{0_2}=(24.0\pm0.4)\times 10^3\;k_BT/\mu m^4$.(D) Representation of
the cylindrical-torus shell proposed, with principal curvatures $C_L
= \frac{1}{R_L}$ and $C_T=\frac{1}{R_T}$. (E) Illustration of a RBC
membrane area delimited by a circumference of radius $R
\sim2\;\mu$m.}\label{figure4a_4e}
\end{figure}

The confining potential $\gamma$ of the fluctuation spectrum of equation
\ref{espectro1} is a term that prevents a planar surface to move
away from its equilibrium position (for $q=0$ mode). For a planar
membrane close to a substrate, if the membrane/substrate interaction
can be modeled by a harmonic potential, then the confining potential
is a measure of its spring constant per unit area \cite{safranbook}.
In the RBC composite membrane model proposed by \emph{Auth et al.}
\cite{safran2007}, the connection of the spectrin-cytoskeleton to
the lipid bilayer is modeled by a confining potential, such that the
lipid bilayer remains sparsely attached to the cytoskeleton by
entropic springs and kept at an average distance from it.
Differently, in closed-shell membrane models the shell size is
limited by the membrane total finite size and the confining
potential is associated to the restoring force due to an area
expansion/compression resultant of the shell breathing mode ($q=0$)
\cite{popescu2010}. In figure \ref{figure4a_4e}(A) and (B) our measurements for $\gamma_1$
and $\gamma_2$ are presented, respectively. The values are within
the range of $[5-50]\times10^3\;k_BT/\mu$m$^4=[20-200]\times10^6\;J/m^4$
(assuming $T$ as the bath temperature). The empty symbols represent the data before correction for lateral movement and the full ones represent the data after correction. The grey area is the standard
deviation of the mean value obtained with the ten final runs of the AG.

By analyzing the data we propose that two types of confining
potential are at work: one related to the cell local curvature and
another related to the cell overall shell shape, named here as local and
global potentials, respectively. In the RBC spherical shell model
proposed by \emph{Park et al.}\cite{popescu2010} the cell is treated
as a spherical shell and the confining potential is associated to
the restoring force that appears due to spatially uniform radius
displacements ($q=0$ breathing mode). For this deformation, the
simple surface energy \cite{boal},
\begin{eqnarray}\label{ka2b}
f_t = \frac{KA}{2}\bigg(\frac{\Delta A}{A}\bigg)^2
\end{eqnarray}
can be considered, where $KA$ is the area compressibility modulus,
$A$ is the surface area ($A=4\pi R^2$, for a spherical shell of
radius $R$) and $\Delta A$ is the area increase/descrease. If the
shell has an equilibrium radius $R_0$, a radial displacement $u$
costs an energy,
\begin{eqnarray}\label{ka3}
f_t = 2KA\frac{u^2}{R_0^2}= 2KA\frac{2}{R_0^2}\frac{u^2}{2}=\gamma
\frac{u^2}{2},
\end{eqnarray}
neglecting $u^4$ terms. For general surface shapes, with principal
curvatures $C_1$ and $C_2$,
\begin{eqnarray}\label{ka4}
f_t = 2KA\bigg(C_1^2+C_2^2\bigg)\frac{u^2}{2}= 4KA(2\mathcal{H}^2 -
K)\frac{u^2}{2},
\end{eqnarray}
which defines, $\gamma = 4KA(2\mathcal{H}^2 - K),$ where
$\mathcal{H}$ is the surface mean curvature and $K$ is the Gaussian
curvature. In figure \ref{figure4a_4e}(C) the plots of
$\gamma$ \emph{versus} $4(2\mathcal{H}^2 - K)$ for each membrane are
presented. The linear equation $\gamma_{1/2} = KA_{1/2}\times
4(2\mathcal{H}_{1/2}^2 - K_{1/2}) + \gamma_{0_{1/2}}$ fits
reasonably well both data and returns similar values for the compressibility
modulus $KA_1=(1.6\pm0.1)\times 10^3\;k_BT/\mu m^2\sim 7\;\mu
N/m$ and $KA_2=(2.0\pm0.3)\times 10^3\;k_BT/\mu m^2\sim 8\;\mu
N/m$ (assuming $T$ as the bath temperature). Those values are of the order of half the ones reported by Park \emph{et al.} in
\cite{popescu2010}, where they found a value of $\sim 18\;\mu N/m$
for whole discocyte RBCs. From the linear fit it becomes clear the
existence of a potential term $\gamma_0$, which is proposed here to
be a global potential responsible to keep the overall shape of the
cell. The values of $\gamma_0$ found for the upper and lower
membranes are $\gamma_{0_1}\sim 10\times 10^3\;k_BT/\mu m^4$ and
$\gamma_{0_2}\sim 24\times 10^3\;k_BT/\mu m^4$, respectively. It is
important to mention that for the upper membrane fitting all RBC region for $\gamma_1$ was used,
while only the more central region ($\rho\leq1\;\mu$m) of the cell,
far from the adhesion region, was used for $\gamma_2$.

To model the global potential $\gamma_{01}$ for the upper membrane a more
realistic description of the red cell shape is used. The cell is
treated as a cylindrical-torus shell (figure \ref{figure4a_4e}(D)),
such that its two curvatures are ($1/R_T$) and ($1/R_L$), with a
potential (equation \ref{ka4}),
\begin{eqnarray}\label{gamma01}
\gamma_{0_1} = 2KA_1\bigg(\frac{1}{R_L^2}+\frac{1}{R_T^2}\bigg).
\end{eqnarray}
Considering $R_T$ as the RBC radius measured experimentally
$R_{rbc}\sim4\;\mu$m, such that $1/R_T\sim0.25\;\mu$m$^{-1}$, and
using the value for $KA_1$ previously measured, one obtains that
$\gamma_{0_1}\sim10\times10^3\;k_BT/\mu m^4$ coincides with equation
\ref{gamma01} if $R_L\simeq0.6\;\mu$m. From figure
\ref{figure4a_4e}(D), a good approximation for the lateral radius
$R_L$ is given by half of the cell maximum thickness $H_{max}/2\sim1\;\mu$m. Finally, it is
noteworthy to mention that the main contribution to $\gamma_1$ comes
from the cylindrical-torus term and not from the local curvature term
$2\mathcal{H}^2 - K$ of the upper membrane, indicating that the
restriction to fluctuation is mostly due the cylindrical envelope,
rather than due the sparse connections between the cytoskeleton and
the lipid bilayer, as proposed by \emph{Gov et al.} \cite{gov2003}.

In order to model the global confining potential $\gamma_{0_2}$ the
lower membrane is treated as a spherical cap of radius $R_0$ and
height $h_0$ able to deform in height but not in radius. This
assumption takes into account that the lower membrane is attached to
the substrate and thus fixed in a circumference of radius $R$, as
depicted in figure \ref{figure4a_4e}(E). For the cap surface area
$A_{0}=2\pi R_0h_0$, a deformation $u$ in the cap's height costs an
energy (equation \ref{ka2b}),
\begin{eqnarray}\label{gamma2_1}
f_t = \frac{KA}{2}\bigg(\frac{\Delta
A}{A}\bigg)^2=\frac{KA}{2}\frac{u^2}{h_0^2}\equiv\gamma_2\frac{u^2}{2},
\end{eqnarray}
which defines the global confining potential,
\begin{eqnarray}\label{gamma2_2}
\gamma_{0_2}\equiv\frac{KA}{h_0^2}.
\end{eqnarray}
Using the experimental values $KA_2 \sim 2\times 10^3\;k_BT/\mu m^2$
and $\gamma_{0_2}\sim24\times 10^3\;k_BT/\mu m^4$ we have $h_0\sim0.3\;\mu$m. Observing
the height profiles of figure \ref{figure2a_2c}(C) it can be seen
that $h_0\sim0.4\;\mu$m, which is close to the estimated
value above as well, supporting the global shape confinement proposed
for the lower membrane. As mentioned before, the linear potential
$\gamma_{2} = KA_{2}\times 4(2\mathcal{H}_{2}^2 - K_{2}) +
\gamma_{0_{2}}$ is a good approximation for points of the membrane
far from the adhesion sites ($\rho \leq 1$), where the membrane
interacts weakly with the substrate. If this potential is
subtracted from the total potential $\gamma_2$, a residual
$\gamma_R$ is obtained. It is suggested here that this residual
potential is related to the substrate presence and thus maximum in
the region of membrane adhesion. Indeed, the
plot of $\gamma_R$ versus the distance of the lower
membrane to the substrate ($h_2$)(plot not shown) is approximately linear and has a maximum value at $h_2=0$
(where the membrane touches the glass coverslip) and is null when $h_2$ achieves its maximum value. Despite that, there
are still many questions regarding this extra potential and more
experiments are necessary for the proposal of a model.

\subsection{Bending modulus $k_c$}

\begin{figure}[t!]
\centering{\includegraphics[width=35pc]{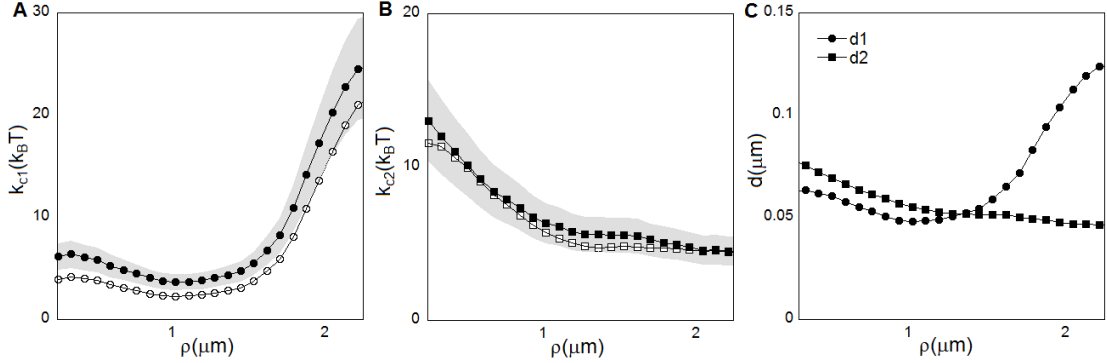}}
\caption{Bending modulus for each cell membrane: (A) $k_{c1}$ and (B) $k_{c2}$. The unfilled
symbols represent the non-corrected data and the filled symbols
represent the corrected ones. The grey area is the standard
deviation of the mean value obtained with the ten final runs of the AG. (C) Thickness profiles of the cell membranes
(bilayer$+$cytoskeleton). The values are within the range of
$(0.050 - 0.125)\;\mu$m, in accordance with compression experiments
\cite{ken}.}\label{figure5a_5c}
\end{figure}

The results for the bending modulus are presented in figure
\ref{figure5a_5c} (A) and (B), for upper and lower membranes, respectively. The empty symbols represent the data before correction for lateral movement and the full ones represent the data after correction. The grey area is the standard
deviation of the mean value obtained with the ten final runs of the AG. According to
\emph{Gov et al.} \cite{gov2003} $k_c$ is assumed approximately
constant along the cell radius. The same is done by \emph{Park et
al.}, who uses in his spherical shell model \cite{popescu2010} an
unique value of $k_c$ for the whole cell. Surprisingly, our data suggests that $k_{c1}$ and $k_{c2}$ vary throughout the cell
membranes within the limits of $[5-30]\;k_BT$. The bending modulus $k_{c1}$ for example has a value of $\sim6\;k_BT$ in the middle of the cell and increases until it achieves a value of $\sim25\;k_BT$ for $\rho\sim 2\;\mu$m. This range comprises the different values of $k_c$ measured for the
whole cell using others techniques \cite{popescu2010,cicuta2009}. Moreover, the observed increase of $k_{c1}$ with the cell radius might be the explanation why techniques that measure membrane fluctuations at the cell edge observe higher values for $k_c$ than others that average the fluctuations over the cell area \cite{popescu2010,cicuta2009}.

A possible explanation to $k_c$ alteration is that the RBC membranes thickness
vary with position, since $KA_1$ and $KA_2$ remain practically
constant along the cell. A relation between $k_c$ and the thickness
$d$ of the membrane is given by \cite{boal},
\begin{eqnarray}\label{boal_1}
k_c= \frac{KAd^2}{\alpha},
\end{eqnarray}
where $\alpha$ depends on the membrane geometry. We propose that each
RBC membrane is composed by two thin elastic sheets (lipid bilayer
and cytoskeleton) locally separated by a distance $d$ (due to
specific proteins) that can slowly vary along the cell radius, with compressibility modulus $KA_m$ for the lipid bilayer and $KA_c$ for the cytoskeleton. Following \emph{Evans et al.} \cite{evans} we have that,
\begin{eqnarray}\label{kc1}
k_c = KA_{eff}d^2,
\end{eqnarray}
where $KA_{eff}$ is an effective compressibility modulus given by,
\begin{eqnarray}\label{kc2}
KA_{eff}=\frac{KA_mKA_c}{KA_m+KA_c}.
\end{eqnarray}
Using as $KA_{eff}$ the values $KA_1$ and $KA_2$ previously determined we obtain the thickness $d_1$ and $d_2$ of the upper and lower membranes, respectively, as shown in
figure \ref{figure5a_5c}(C). The dots represent
the upper membrane thickness ($d_1$) and the squares represent the
lower membrane thickness ($d_2$). The thickness values correspond to
the bilayer$+$cytoskeleton width and are within the range of ($0.050
- 0.125)\;\mu$m. Notably, those values are in agreement with
compression experiments performed using a biomembrane force probe
and optical interferometry \cite{ken}. Analyzing the upper membrane
thickness profile it can be seen that $d_1$ significantly increases
for radius positions beyond $1.50\;\mu$m, a behavior which was
theoretically predicted by \emph{Y. Fung and P. Tong} \cite{tong}.
Additionally, the lower membrane thickness profile indicates that
the membranes have approximately the same thickness value in the
center of the cell, and decreases as the membrane gets closer to the
substrate. The interaction membrane$/$glass substrate may be
inducing a compression on the lower membrane.

\subsection{Surface tension $\sigma$}

\begin{figure}[t!]
\centering{\includegraphics[width=28pc]{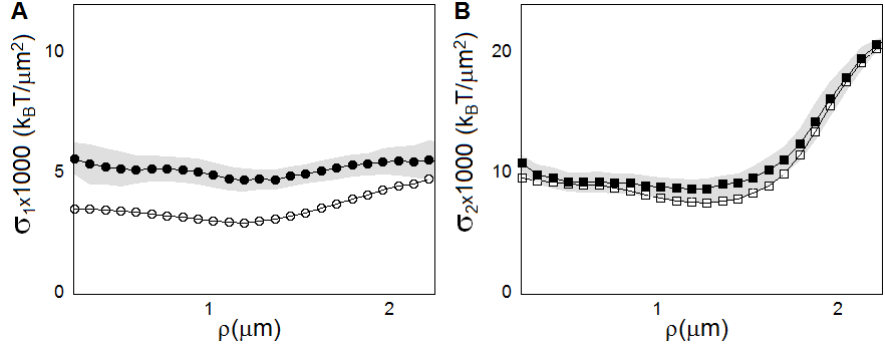}}
\caption{Surface tension for each cell membrane: (A) $\sigma_{1}$ and (B) $\sigma_{2}$. The unfilled
symbols represent the non-corrected data and the filled symbols
represent the corrected ones. The grey area is the standard
deviation of the mean value obtained with the ten final runs of the AG.}\label{figure6a_6b}
\end{figure}

In figure \ref{figure6a_6b}(A) and (B) the results for the surface
tension $\sigma_1$ and $\sigma_2$ are presented, respectively. The empty symbols represent the data before correction for lateral movement and the full ones represent the data after correction. The grey area is the standard
deviation of the mean value obtained with the ten final runs of the AG. The
range of values $[20-80]\;\mu N/m$ (considering $T$ as the bath
temperature), is at least tenfold higher than the ones reported by
others authors \cite{popescu2008,cicuta2009}.
One possible explanation for that discrepancy is that the values are a measure of
a mechanical tension and not a surface tension, as originally
proposed by \emph{Gov et al.} \cite{gov2003}. According to \emph{Farago}
\cite{farago2011}, the physical meaning of the term $\sigma$ in the
spectrum of equation \ref{espectro1} is actually a mechanical tension.
\emph{Farago} \cite{farago2011} defines the term $\sigma$ as the lateral
force per unit length exerted on the boundaries of a membrane when
one attaches it to a frame. This idea is illustrated in figure
\ref{figure4a_4e}(E), for the case of a red cell, where the membrane
area adhered to the substrate is delimited by a circumference of
radius $R \cong 2\;\mu$m. If this is the case, the mechanical
tension can be estimated as the adhesion force $F_a$ by which the
membrane is being pulled by the substrate, divided by the
circumference length of the adhered region,
\begin{eqnarray}\label{ka5}
\sigma\sim\frac{F_a}{2\pi R}.
\end{eqnarray}
Considering $\sigma_1= 7\times10^3k_BT/\mu$m$^2 \equiv
28\;\mu$N$/$m (with $T$ the bath temperature), the adhesion force
obtained for the upper membrane is $\sim350\;$pN.  For $\sigma_2$,
the estimated force is approximately $F_a\sim500\;$pN. Micropipette measurements \cite{gingell} of the adhesion force of
red cells to common glass (hydrophilic) are between $500$ to
$1600\;$pN.

\section{Conclusions}

In this paper the formulation of Defocusing Microscopy (DM) is
exposed, together with its application to RBCs. Using the presented
methods we can obtain total $3D$ total images of RBCs, volume, tridimensional thickness profile and determine the height profiles
of each cell surface-membranes (bilayer$+$cytoskeleton), the upper
one free to fluctuate and the lower one adhered to the substrate.
Furthermore, by measuring DM contrast fluctuation and assuming a plane
membrane model, the membrane height fluctuations along the cell radius of each
surface-membrane can be extracted, separately, an unique feature of DM. The radial behavior of the elastic moduli of each
surface-membranes are also obtained, allowing the discussion of
physical aspects of the cell membrane. Finally, the methods
presented here allow the extraction of biomechanical characteristics
of red blood cells and thus can be used as tools to study several
pathologies related to RBCs alteration.

\section{Acknowledgments}

\ack{The authors acknowledge the support from the Brazilian agencies
FAPEMIG, CAPES, CNPq, PRONEX-FACEPE, INFCx and declare no conflicts of
interest. The authors also acknowledge B E de Faria and L Cescato for their help with the gratings experiments.}

\section*{Appendix}

\appendix

\subsection*{Appendix A: Defocusing microscopy}
\setcounter{section}{1}
A monochromatic light electric field $E(\vec{r})$ with propagation
direction along the $z$ axis is considered for propagation. Fourier Optics \cite{goodman} is used, in a way that,

\begin{eqnarray}\label{fouriercampo2}
E(\vec{\rho})=\frac{1}{(2\pi)^2}\displaystyle{\int}
A(\vec{q})\:e^{i\vec{q}\cdot\vec{\rho}}\:d\vec{q}.
\end{eqnarray}
with its inverse transform representing the angular spectrum
$A(\vec{q},z)$ of the electric field,
\begin{eqnarray}\label{espectroangular}
A(\vec{q})=\displaystyle{\int}
E(\vec{\rho})\:e^{-i\vec{q}\cdot\vec{\rho}}\:\vec{\rho}.
\end{eqnarray}
The angular spectrum $A$ is propagated through the phase object,
lenses and free regions \cite{bira2003}. Using the paraxial approximation, the calculation returns
the final electric field at image plane IP,
\begin{eqnarray}\label{eplanoi}
E(\vec{\rho},z_f) = \displaystyle
B\:e^{i\alpha(\vec{\rho})}{(2\pi)^2}\int
A_{0}(\vec{q})\:e^{i\frac{z_f}{2k}q^2}\:e^{i\vec{q}\cdot\vec{\rho}}\:d\vec{q},
\end{eqnarray}
where B is a constant, $\alpha(\vec{\rho})$ is a phase factor and $A_0(\vec{q})$
is the angular spectrum immediately after crossing the specimen
($z=0$), $k=n_{ob}k_0$, with $n_{ob}$ the objective immersion medium
refractive index and $k_0$ the vacuum light wavenumber. As it can be
seen, a bright-field defocused microscope is similar to a
phase-contrast microscope, but with a phase difference between the
diffracted and nondiffracted light of $\Delta
\phi=\frac{z_fq^2}{2k}$.

Considering the phase object as a specimen composed of a single thin
interface with height profile $h(\vec{\rho})$ and refractive index
$n + \Delta n$, immersed in a medium of refractive index $n$, when
an incident plane wave with field $E_0$ crosses the object, it is diffracted \cite{wolf} such that,
\begin{eqnarray}\label{campofase1}
E(\vec{\rho})=E_0\:e^{i\Delta\varphi(\vec{\rho})}=E_0\:e ^{-i\Delta
nk_0h (\vec{\rho})}.
\end{eqnarray}
In the limit of $\Delta \varphi(\vec{\rho})<<1$, $E_0(\vec{\rho},z)\simeq
E_0[1+i\varphi(\vec{\rho})]=E[1-i\Delta nk_0h (\vec{\rho})]$ and using
the Fourier decomposition for the height profile
$h(\vec{\rho})=\frac{1}{\sqrt{S}}\sum_{\vec{q'}}h(\vec{q'})\sin(\vec{q'}\cdot\vec{\rho})$,
\begin{eqnarray}\label{campofasefinal}
E(\vec{\rho})=E\Bigg[1-\frac{i\Delta
nk_{0}}{\sqrt{S}}\sum_{\vec{q'}}h(\vec{q'})\sin(\vec{q'}\cdot\vec{\rho})\Bigg].
\end{eqnarray}
Substituting it into equation \ref{fouriercampo2},
\begin{eqnarray}\label{espectrofasefinal}
A_{0}(\vec{q})=(2\pi)^{2}E_{0}\Bigg[\delta(\vec{q})+ \frac{\Delta
nk_{0}}{2\sqrt{S}}\sum_{\vec{q'}}h(\vec{q'})\:\delta(\vec{q}+\vec{q'})-
\frac{\Delta
nk_{0}}{2\sqrt{S}}\sum_{\vec{q'}}h(\vec{q'})\:\delta(\vec{q}-\vec{q'})\Bigg],\nonumber\\
\end{eqnarray}
which is the angular spectrum after the object. Using this $A_{0}$ in equation
 \ref{eplanoi} and defining the image contrast as
$C(\vec{\rho})=\frac{I(\vec{\rho})-I_0}{I_0}$, where $I(\vec{\rho})$
is the intensity of the object image and $I_0$ is the background
intensity, for first-order diffraction, the DM contrast is given by,
\begin{eqnarray}\label{contrasteultimo}
C(\vec{\rho})=-\frac{2\Delta
nk_{0}}{\sqrt{S}}\sum_{\vec{q}}\Bigg[h(\vec{q})\sin(\vec{q}\cdot\vec{\rho})\sin\Bigg(\frac{z_f}{2k}q^2\Bigg)\Bigg].
\end{eqnarray}

For red blood cells the light electric field passes
through both surface-membranes as in the model of
figure \ref{figure1a_1e}(B). By
assuming weakly diffracting objects, the incident
light is diffracted by one or the other surface, with a negligible
contribution due to diffraction by both surfaces. In this case the defocused electric field can be written as
\cite{bira2003,coelho2005,coelho2007,leo,giuseppe},
\begin{eqnarray}\label{ap3g}
E(\vec{\rho},z_f)=\frac{1}{(2\pi)^2}\displaystyle{\int}
[A_1(\vec{q})\:e^{i\vec{q}\cdot\vec{\rho}}\:e^{i(\frac{(z_f-h_1(\vec{\rho}))q^2}{2k})}+\nonumber\\
A_2(\vec{q})\:e^{i\vec{q}\cdot\vec{\rho}}\:e^{i(\frac{(z_f-h_2(\vec{\rho}))q^2}{2k})}]d\vec{q} - E_0\nonumber\\
\end{eqnarray}
where $h_1(\vec{\rho})$, $h_2(\vec{\rho})$ are the height profiles
for upper and lower RBC surface-membranes (bilayer+cytoskeleton),
respectively, and the terms,
\begin{eqnarray}\label{ap4g}
A_{1}(\vec{q})=\displaystyle{\int}E_{0}\:e^{-i\vec{q}\cdot\vec{\rho}}\:e^{i{\Delta
nk_0h_1(\vec{\rho})}}\:d\vec{\rho}\nonumber\\
A_{2}(\vec{q})=\displaystyle{\int}E_{0}\:e^{-i\vec{q}\cdot\vec{\rho}}\:e^{-i{\Delta
n k_0h_2(\vec{\rho})}}\:d\vec{\rho}
\end{eqnarray}
are the Angular Spectra of the electric field diffracted by the
membranes. Additionally, $\Delta n$ is the refractive index
difference between the RBC and its surrounding medium. Considering equations
 \ref{ap3g} and \ref{ap4g}, for first-order diffraction the DM
contrast for a red cell is,
\begin{eqnarray}\label{ap2g}
C(\vec{\rho})=\frac{2\Delta nk_{0}}{\sqrt{S}}\Bigg\{\sum_{\vec{q}}
\Bigg[h_2(\vec{q})\sin\Bigg(\frac{(z_f-h_2(\vec{\rho}))q^2}{2k}\Bigg) \nonumber\\
-h_1(\vec{q})\sin\Bigg(\frac{(z_f-h_1(\vec{\rho}))q^2}{2k}\Bigg)\Bigg]\sin(\vec{q}\cdot\vec{\rho})\Bigg\},
\end{eqnarray}
where $S$ is the RBC surface area.

\subsection*{Appendix B: DM control experiments}

In order to test the validity of the expressions above for DM, control experiments were
performed using phase gratings with known amplitudes. The gratings height profiles were determined using Atomic Force Microscopy (AFM). The grating DM contrast measurements were conducted with a $100X$ oil immersion objective and using red filtered light of $\lambda=(0.660\pm0.010)\;\mu$m. The images were captured with a CCD camera at $15$fps and the defocusing distance was controlled by a
piezoelectric nanoposition stage. For details see Materials and methods section.

If a phase object with periodic structure is visualized through a
defocused microscope its image contrast is repeatedly seen for a
range of defocus distances. For a sinusoidal phase grating with
height $h(x)=h\sin(q_gx)$, where the grating's wavenumber is $q_g=\frac{2\pi}{\Lambda}$ and $\Lambda$ is the groove spacing, using equation \ref{contrasteultimo} (since $z\gg h$) the contrast (first order diffraction) along the $x$ axis is given by,
\begin{eqnarray}\label{contrasterede}
C(x,z_f)= 2\Delta nk_{0}h \sin(q_gx)
\sin\bigg(\frac{z_f{q_g}^2}{2n_{ob}k_0}\bigg),
\end{eqnarray}
where $k_0$ is the incident wavenumber, $n_{ob}$ is the objective
immersion medium refractive index and $\Delta n$ is the refractive
index difference between the grating and its surrounding medium. The
contrast as shown in equation \ref{contrasterede} is a combination of two sine terms, so that when the $z$
axis is scanned the grating self-image fades and reappears until the
incident light coherence is lost.

From the maximum amplitude of a grating contrast ($C_{max}=(2\Delta nk_{0}h)$), obtained when $\frac{z_f{q_g}^2}{2n_{ob}k_0}=\frac{\pi}{2}$, the grating's height $h$ can be determined if $\Delta n$ is known. Also, from the contrast periodicity in the $x$ axis, given by $\sin(q_gx)$, the grating's groove spacing can be found, since $q_g = \frac{2\pi}{\Lambda}$. Finally, from the contrast periodicity in the $z$ axis, given by $\sin\bigg(\frac{z_f{q_g}^2}{2n_{ob}k_0}\bigg)$, it can be confirmed that the appropriate light wavenumber to be used in the defocusing contrast term is $k=k_0n_{ob}$. Here we use both the square contrast and contrast correlation in $x$ to obtain the grating's parameters, since it provides a more accurate determination of the parameters involved due to a better averaging along the grating. The results are then compared with the ones measured using different techniques.

\begin{figure}[\center\here]
\centering\includegraphics[width=14cm]{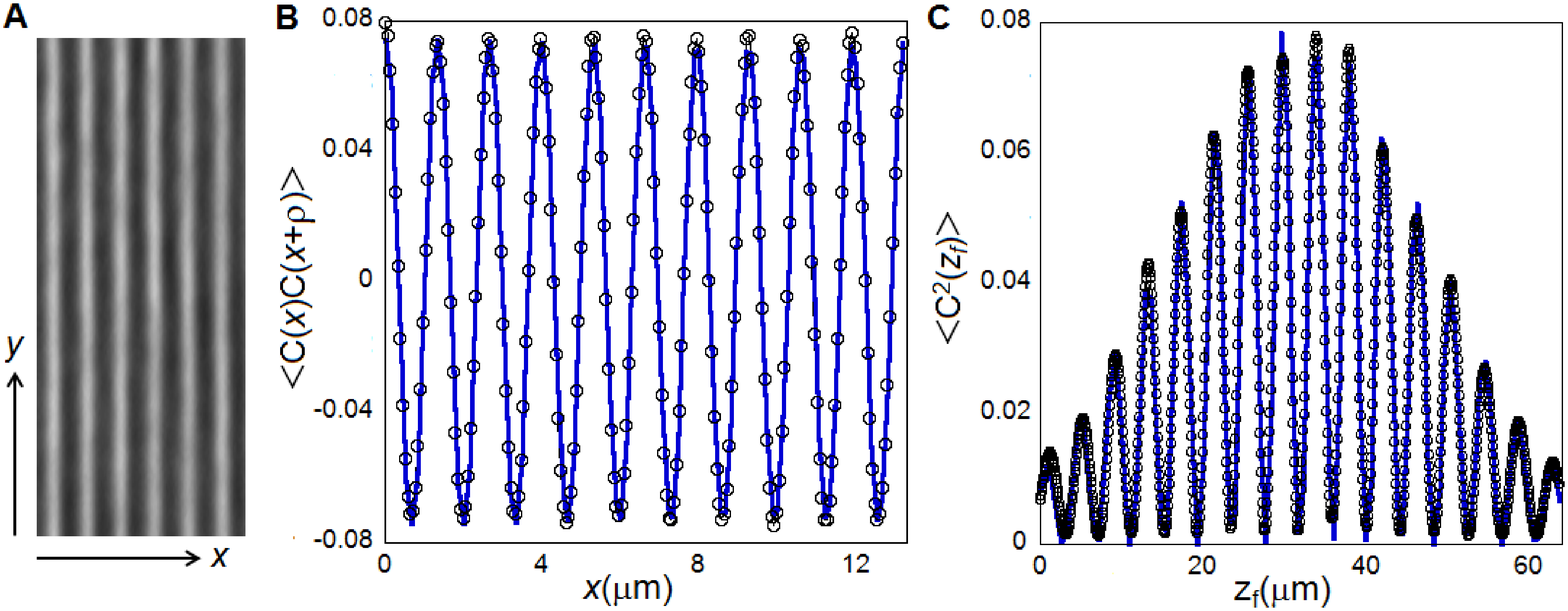}
\caption{(A)Image of a phase grating contrast at its maximum amplitude position. (B) Correlation of contrast $<C(x)C(x+\rho)>$ along a horizontal line ($x$ axis).(C)Average square contrast $<C^2(z_f)>$ along the $z$ axis. The grating contrast along $z_f$ varies periodically with a periodicity of $\lambda_{DM}$.}\label{figureB1a_B1c}
\end{figure}

An image of a phase grating contrast at its maximum amplitude position is shown in figure \ref{figureB1a_B1c}(A). In (B) the correlation contrast $<C(x')C(x'+ x)>$ along a horizontal line ($x$ axis) of (A) is presented. The contrast correlation in $x$ is given by,
\begin{eqnarray}\label{cor_x_rede}
<C(x')C(x' + x)>= (\sqrt{2}\Delta nk_{0}h)^2\cos(q_gx),
\end{eqnarray}
which can be fitted to the data of plot (B), letting $(\sqrt{2}\Delta nk_{0}h)^2$ and $\Lambda$ as free parameters. The value for $\Lambda$ with the error bar as returned by the fit is $\Lambda=(1.3219\pm0.0001)\;\mu$m, which is in agreement with AFM measurement ${\Lambda_{AFM}} = (1.33 \pm 0.05)\;\mu$m.

To determine the grating's height it is necessary to know the grating's refractive index, which can also be obtained experimentally using DM contrast measurements. When the grating is immersed in air, the maximum contrast observed for the grating is $C_{air}=2(n_g-n_{air})k_0h$. However, if the grating is immersed in a different medium, with refractive index $n_{m}$, then the maximum contrast becomes $C_{m}=2(n_g-n_{m})k_0h$, such that,
\begin{eqnarray}\label{indice_refracao1}
\frac{C_{air}}{C_m}=\frac{n_g-n_{air}}{n_g-n_{m}}\;\;\;\mbox{and}\;\;\;\mbox{thus,}\;\;\;n_g=\frac{C_{air}n_{m}-C_{m}}{C_{air}-C_m}.
\end{eqnarray}
In order to determine the grating refractive index we have covered it with an oil of known refractive index ($n_{oil}=1.51\pm0.01$). In that configuration we obtained a maximum contrast of $C_{m}=0.063$. In air, the maximum contrast for the same grating was $C_{air}=0.31$, such that the refractive index for the analyzed grating was found to be $n_g=1.64\pm0.02$. This value is $4\%$ higher than the one measured using the prism coupler technique.

In figure \ref{figureB1a_B1c} (C) the average square contrast $<C^2(z_f)>=\frac{1}{A}\int\int C^2(x,z_f)dxdy$ as a function of $z_f$ axis is presented since this is an average over the entire image and provides better statistics. The plot was obtained by using the Histogram tool of Image J (Rasband WS. ImageJ, U.S. National Institutes of Health, Bethesda, Maryland, USA,
imagej.nih.gov/ij/, 1997–2012) \cite{image} for the entire image of each frame, where the standard deviation divided by the mean intensity is equal to $\sqrt{<C^2(z_f)>}$. By squaring it we obtain $<C^2(z_f)>$. The envelope function of the incident light can be approximated by a gaussian function $\:e^{\frac{-(\lambda-\lambda_0)^2}{2\Delta{\lambda}^2}}$, where $\lambda_0 = 0.660\;\mu$m and $\Delta \lambda=0.010\;\mu$m is the filter width, such that,
\begin{eqnarray}\label{c_std}
<C^2(z_f)>=(\sqrt{2}\Delta nhk_0)^2\sin^2\bigg(\frac{(z_f-z_0)^2{q_g}}{2n_{ob}k_0}\bigg)\:e^{\frac{-(z_f-z_0)^2}{2\xi_z^2}},
\end{eqnarray}
where $\xi_z=\frac{\Lambda^2n_{ob}}{2\pi\Delta\lambda}$ is the axial coherence length in this case. The function $<C^2(z_f)>$ varies periodically as a function of the distance $z_f$ with
the DM wavelength $\lambda_{DM}$,
\begin{eqnarray}\label{lambda_dm}
\frac{2\pi}{\lambda_{DM}}=\frac{q_g^2}{2n_{ob}k_0}=\frac{2\pi\lambda_0}{2\Lambda^2n_{ob}},
\end{eqnarray}
such that $\lambda_{DM} = n_{ob}\frac{2\Lambda^2}{\lambda_0}$, which can be related to the Talbot length $\lambda_T = \frac{2\Lambda^2}{\lambda_0}$ \cite{rayleigh}, then
\begin{eqnarray}\label{lambda_dm2}
\lambda_{DM}=n_{ob}\Lambda_T.
\end{eqnarray}
Equation \ref{c_std} was used to fit the data of $<C^2(z_f)>$, leaving the terms $(\sqrt{2}\Delta nhk_0)^2$, $\lambda_{DM}$ and $\xi_z$ and $z_0$ as free parameters. The best fit is presented in figure \ref{figureB1a_B1c}(C) as the blue line and returned the values $0.0793\pm0.0003$, $(8.274\pm0.0002)\;\mu$m, $(16.02\pm0.07)\;\mu$m and $(31.683\pm0.003)\;\mu$m, respectively. Using the value for $(\sqrt{2}\Delta nhk_0)^2$, $\lambda_0=0.660\;\mu$m and $\Delta n=0.064$ (determined previously) we get the grating's height $h=0.033\pm0.001\;\mu$m, which compares well with the value $h_{AFM}=0.0361\pm0.0002\;\mu$m returned by AFM measurements. From the fit we also find that $\lambda_{DM}=(8.274\pm0.002)\;\mu$m together with the Talbot length $\lambda_T = (5.30\pm0.08)\;\mu$m, which results from equation \ref{lambda_dm2} the value $n_{ob}=1.56\pm0.02$, slightly higher than the manufacturer's value of $1.51$. This confirms that $k=n_{ob}k_0$, as predicted by our model of defocusing, as the proper wavenumber $k$ to be used. In the case of a dry objective the correct wavenumber is $k=k_0$. Additionally, if
the recording video-camera is the element displaced to cause image
defocusing, then the phase shift is scaled by other parameters and
one can relate the displacement of the video-camera to the
displacement of the objective to produce the same amount of
defocusing. In this case, it is obtained that
$z_{camera}=M^2z_{objective}$, where M is the magnification of the
objective. Finally, the value for $\xi_z=(16.02\pm0.07)\;\mu$m gives the axial coherence length. From the expression for $\xi_z$ we obtain that $\Delta\lambda\sim0.027\;\mu$m, larger than the filter width $\Delta \lambda=0.010\;\mu$m, which suggests that additional effects are occurring.

\subsection*{Appendix C: Genetic Algorithm}

A GA is a search algorithm based on the mechanisms of Darwinian
evolution that uses random mutation, crossover, and selection
operators to breed better solutions from an originally random
starting population \cite{koza}. GAs have been widely used to solve complex problems in different areas of knowledge like engineering, management and medicine. In this paper a GA was implemented and employed to obtain the power spectrums $<|{u_1(\vec{q})}|^2>$ and $<|{u_2(\vec{q})}|^2>$ using
equation \ref{espectro1} as fitting equation. Preliminary tests with five of the major benchmarking functions for GA testing presented in \cite{AG_2} were used to evaluate its performance. The five selected functions have known global optimal and different degrees of complexity: F$1$ is smooth unimodal and strongly convex, F$2$ has a very narrow ridge with sharp tips, F$5$ has many local optima, F$7$ has a wide search space and a large number of local minima and F$8$ is nonlinear and multimodal. GAs have difficulties to converge close to the minimum of such functions, because due to their characteristics, the probability of making progress decreases rapidly as the global optimal is approached. The tests consisted in $100$ executions for each function with $90000$ function evaluation per execution. The GA converged in all tests achieving high repeatability with small standard deviation of the solutions, showing that it is capable of solving complex problems. The searches were started with
a random population of $x$ individuals $20$ fold the number of
variables of the fitting equation. For equation \ref{deltacrbc1} each
individual described an elastic parameter, such that the number of
variables was $150$ and an initial population of $3000$ individuals
was used. In this case, a search space was defined for each elastic
parameter: $k_c=[0,100]k_BT$, $\gamma=[0,50000]k_BT/\mu$m$^4$ and
$\sigma=[0,500000]k_BT/\mu$m$^2$ where $k_B$ is the Boltzmann
constant and T temperature in Kelvin ($T = 299\;K$ for the
experiments). Those limits were continuously decreased as the
algorithm zone of convergence was delimited. Giving a random
starting population, the GA iteratively performs evolution
operations to breed a new generation of individuals, until some stop
criterion is satisfied \cite{koza}. More specifically, in each
generation the individual's fitness is assessed by calculating $\Psi^2$, given by $\Psi^2 = \sum_{i}^{N}(T_i - t_i)^2$,
where $T$ is the target data set ($<\Delta C^2(i)>$), $t$ is the
individual's data set of the generation and $N$ is the number of
variables. Theoretically, the stop criteria should be that $\Psi^2$ reaches the value of zero. However, since the target data
set is an experimental measurement, the intrinsic uncertainties must
be taken into account. Therefore, the final solution was found when
consecutive fitness values returned by best individuals had
not changed by more than $1\%$ of the experimental data variance for
$1000$ generations, starting the count from fitness values
below $10\%$ of the experimental data variance. In order to qualitatively access the robustness of the
algorithm to initial conditions, the GA was executed several times
with random initial population. In each run an average of
$5\times10^{6}$ function evaluations were necessary for the
algorithm to converge. In figure \ref{figure3a_3c}(A) the ten final fitting
results for $<\Delta C^2>$ over $42$ RBC are presented. The AG takes approximately $10$ hours
to converge in computers with $i7$ Intel Core (Intel Company, Santa Clara, CA) processors and $32G$ of RAM memory.

\subsection*{Appendix D: Correction for cell lateral movement}

\begin{figure}[t!]
\centering{\includegraphics[width=26pc]{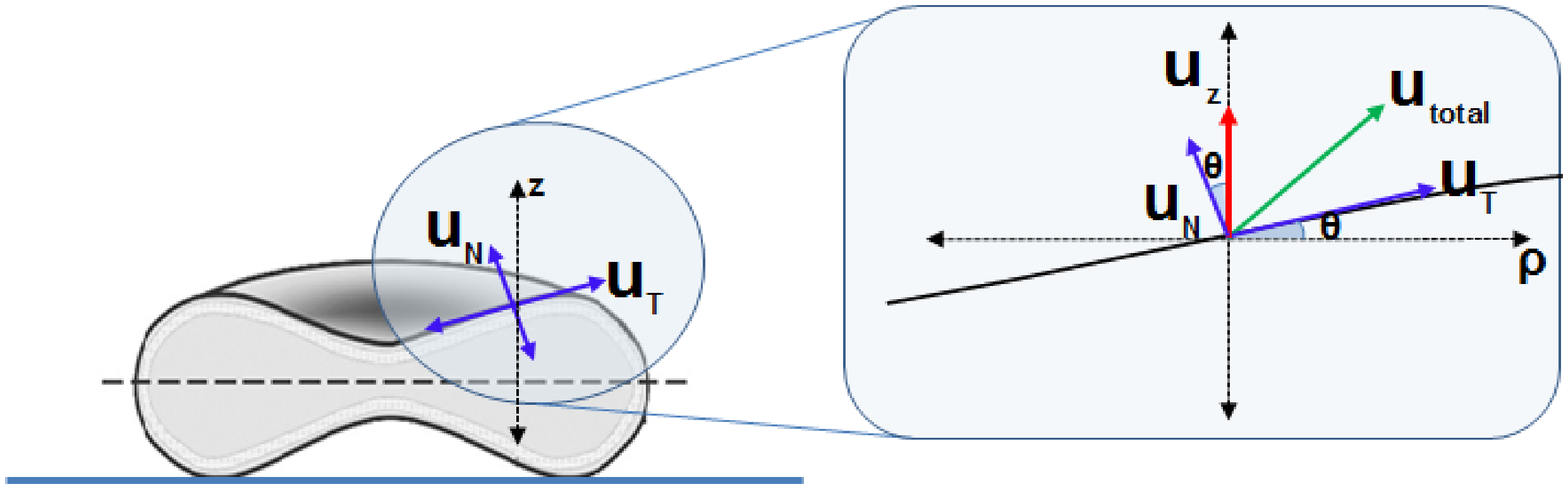}}
\caption{The two components of the cell total fluctuation $u_{total}$: $u_{T}$ along the surface tangent and $u_{N}$ along the normal to the surface. The $z$ axis represents the axis along which the fluctuation is collected in DM technique ($u_z$). To obtain $u_N$ (height fluctuation) from $u_z$, a correction is needed.}\label{figureD1a}
\end{figure}

What is measured with DM is the membrane fluctuation along the $z$
axis ($u_z$). This fluctuation can be decomposed into a normal ($u_N$) and a tangent ($u_T$) components to the membrane surface, as in figure \ref{figureD1a}. Thus, $u_z$ is the $z$ projection of $\vec{u}_{total}=\vec{u}_N + \vec{u}_T$
and since equation \ref{espectro1} is related to the fluctuation $u_N$, $u_N$ has to be extracted from the data of $u_z$. In figure \ref{figureD1a},
$\theta$ is the angle formed between $u_N$ and the $z$ axis, then $<(u_z)^2>
= <(u_N)^2> \cos^2\theta+ <(u_T)^2>\sin^2\theta$, and is also the angle
formed between $u_T$ and the $\rho$ axis, such that $\tan\theta =
\frac{dh}{d\rho}$, where $h$ is the membrane height profile. The
value for $u_z$ is equal $u_N$ when $\theta = 0$ (middle of the
cell), where from figure \ref{figure3a_3c} (B) we see that $u_{z(rms)}\sim 10\;$nm.
Additionally, $u_T$ may be assumed to be caused by the cell shear
movement, $<(u_T)^2> \simeq \frac{k_BT}{\mu}\simeq
5\times10^{-16}m^2$, for a shear modulus $8\times10^{-6}\mu Nm^{-1}$
\cite{popescu2010}, resulting in a $u_{T(rms)}$ value of $22\;$nm.
Measuring the cell border fluctuations (where $u_{N(rms)}\sim 0$)
using DM, it is found that $u_{T(rms)} \sim (16-22)$ nm, in
agreement with the above estimation. Since at the cell center, where
$u_T\sim 0$, the value $u_{N(rms)}\sim 10$nm is obtained, then
$u_{T(rms)}\simeq 2u_{N(rms)}$ and
\begin{eqnarray}\label{correcao5}
<(u_N)^2> \cong \frac{<(u_z)^2>}{4 - 3cos^2\theta}.
\end{eqnarray}

\section*{References}

\begin{thebibliography}{99}

\bibitem{diez2010} Diez-Silva M, Dao M, Han J,Lim C T and Suresh S 2010 Shape and biomechanical characteristics of human red blood cells in health and disease \textit{MRS Bull} {\bf 35} 382

\bibitem{bira2003} Agero U, Monken C H, Ropert C, Gazzinelli R T and Mesquita O N 2003 Cell Surface Fluctuations Studied with Defocusing Microscopy \textit{Phys. Rev.} E {\bf 67} 051904

\bibitem{coelho2005} Neto J C, Agero U, Oliveira D C, Gazzinelli R T and Mesquita O N 2005 Real-time measuraments of surface dynamics on macrophages and the phagocytosis of Leishmania parasites \textit{Exp. Cell Res.} {\bf 303} 207-34

\bibitem {coelho2007} Neto J C, Agero U, Gazzinelli R T and Mesquita O N 2006 Measuring optical and mechanical properties of a living cell with Defocusing Microscopy. \textit{Biophys. J.} {\bf 91} 1108-15

\bibitem {leo} Mesquita L G, Agero U, and Mesquita O N 2006 Defocusing Microscopy: an approach for red blood cell optics \textit{Appl. Phys. Lett.} {\bf 88} 133901

\bibitem{giuseppe} Glionna G, Oliveira C K, Siman L G, Moyses H W, Prado D M U, Monken C H and Mesquita O N 2009 Tomography of fluctuating biological interfaces using defocusing microscopy \textit{Appl. Phys. Lett.} {\bf 94} 193701

\bibitem{paula} Roma P M S, Siman L, Amaral F T, Agero U and Mesquita O N 2014 Total three-dimensional imaging of phase objects using defocusing microscopy: Application to red blood cells \textit{Appl. Phys. Lett.} {\bf 104} 251107

\bibitem{sebastian_jbo} Etcheverry S, Gallardo M J, Solano P, Suwalsky M, Mesquita O N and
Saavedraa C 2012 Real-time study of shape and thermal fluctuations in the echinocyte transformation of human erythrocytes using defocusing microscopy \textit{J. Biomed. Opt.} {\bf 17} 106013

\bibitem{mariajose} Suwalsky M, Belmar J, Villena F, Gallardo M J,
Jemiola-Rzeminska M, Strzalka K 2013 Acetylsalicylic acid (aspirin) and salicylic acid interaction with the
human erythrocyte membrane bilayer induce in vitro changes in
the morphology of erythrocytes \textit{Arch. of Biochem. and Biophys.} {\bf 539} 9–19

\bibitem {zernicke1} Zernike F 1942 Phase-contrast, a new method for microscopic observation of transparent objects, part I \textit{Physica.} {\bf 9} 686-98

\bibitem {nomarski} Nomarski G 1955 Nouveau dispositif pour l'observation en contraste de phase differentiel \textit{J. Phys. Radium.} {\bf 16} 9-11

\bibitem {gureyev97} Gureyev T E and Nugent K A 1997 Rapid quantitative phase imaging using the transport intensity equation \textit{Opt. Commun.} {\bf 133} 339-46

\bibitem {nugent1} Paganin D and Nugent K A 1998 Noninterferometric phase imaging with partially coherent light \textit{Phys. Rev. Lett.} {\bf 80} 2586-89

\bibitem {nugent2} Barone-Nugent E D, Barty A and Nugent K A 2002 Quantitative phase-amplitude microscopy I: optical microscopy. \textit{J. Micros.} {\bf206} 194-206

\bibitem {popescu2008} Popescu G, Park Y, Choia W, Dasaria R R, Feld M S, and Badizadegana K 2008 Imaging red blood cell dynamics by quantitative phase microscopy. \textit{Blood Cells, Mol. Dis.} {\bf 41} 10-6

\bibitem {kononenko} Kononenko V L 2011 Characterization of red blood cells rheological and physiological state using optical
flicker spectroscopy \textit{Advanced Optical Flow Cytometry: Methods
and Disease Diagnoses} ed V. V. Tuchin (Weinheim: Wiley-VCH Verlag GmbH \& Co. KGaA) pp 155-210

\bibitem {pnas2010} Park Y, Best C A, Auth T, Gov N, Safran S, Popescu G, Suresh S and Feld M S 2010 Metabolic remodeling of the human red blood cell membrane \textit{Proc. Natl. Acad. of Sci.} {\bf 107} 1289-94

\bibitem {popescu2010} Park Y, Best C A, Badizadegana K, Dasari R R, Feld M S, Kuriabova T, Henle M L, Levine A J and Popescu G 2010 Measurement of red blood cell mechanics during morphological changes. \textit{Proc. Natl. Acad. of Sci.} {\bf 107} 6731-36

\bibitem {teague} Teague M R 1983 Deterministic phase retrieval: a Green's function solution. \textit{J. Opt. Soc. Am.} {\bf 73} 1434-41

\bibitem{popescunature} Kim T, Zhou R, Mir M, Babacan S D, Carney P S, Goddard L L and Popescu G 2014 White light diffraction tomography of unlabelled live cells \textit{Nature Photon.} {\bf 8} 256–63

\bibitem {wolf} Born M and Wolf E 1999 \textit{Principles of Optics} (Cambrigde: Cambridge University Press)

\bibitem{brochard} Brochard F and Lennon J F 1975 Frequency spectrum of the flicker phenomenon in erythrocytes \textit{Le Journal de Physique} {\bf 11} 1035-47

\bibitem{safran2007} Auth T, Safran S A and Gov N S 2007 Fluctutations of coupled fluid and solid membranes with application to red blood cells \textit{Phy. Rev.} E {\bf 76} 051910

\bibitem{korenstein1991} Levin S, and Korenstein R 1991 Membrane fluctuations in erythrocytes are linked to MgATP-dependent dynamic assembly of the membrane skeleton, \textit{Biophys J.} {\bf 60} 733-37

\bibitem{cicuta2009} Yoon Y, Hong H, Brown A, Kim D C, Kang D J, Lew V L and Cicuta P 2009 Flickering analysis of erythrocyte mechanical properties: dependence on oxygenation Level, cell Shape, and hydration Level \textit{Biophys. J.} {\bf 97} 1606–15

\bibitem{image} Schneider C A, Rasband W S and Eliceiri K W 2012 NIH Image to ImageJ: $25$ years of image analysis \textit{Nat. Methods} {\bf 9} 671-75

\bibitem{rappaz2008} Rappaz B, Barbul A, Emery Y, Korenstein R, Depeursinge C, Magistretti P J and Marquet P 2008 Comparative study of human erytrocytes by digital holographic microscopy, confocal microscopy and impedance volume analyzer \textit{Cytometry} A {\bf 73} 895

\bibitem{rappaz2012} Boss D, Hoffmann A, Rappaz B, Depeursinge C, Magistretti P J, Van de Ville D and Marquet P 2012 Spatially-resolved eigenmode decomposition of red blood cells membrane fluctuations questions the role of ATP in flickering \textit{PLoS One} {\bf 8} 1-10

\bibitem{safranbook} Safran S 1994 \textit{Statistical thermodynamics of surfaces, interfaces, and membranes} (Massachusetts: Addison-Wesley Pub.)

\bibitem{boal} Boal D 2002 \textit{Mechanics of the Cell} (Cambridge: Cambrigde University Press)

\bibitem{gov2003} Gov, N., A. Zilman and S. Safran. 2003. Cytoskeleton confinement and tension of red blood cell Membranes. Phys. Rev. Lett. 90:228101.

\bibitem{evans} Evans E 1974 Bending resistence and chemically induced moments in membrane bilayers \textit{Biophys. J.} {\bf 14} 923-31

\bibitem{ken} Heinrich V, Ritchie K, Mohandas N and Evans E 2001 Elastic thickness compressibilty of the red cell membrane. \textit{Biophys. J.} {\bf 81} 1452-63

\bibitem{tong} Fung Y and Tong P 1968 Theory of sphering of red blood cells \textit{Biophys. J.} {\bf 18} 175-98

\bibitem{farago2011} Farago O 2011 Mechanical surface tension governs membrane thermal fluctuations \textit{Phys. Rev.} E {\bf 84} 051914

\bibitem{gingell} Francis G W, Fisher L R, Gamble R A and Gingell D 1987 Direct measurements of cell detachment force on single cells using a new electromechanical method \textit{J. of Cell Sci.} {\bf 87} 519-23

\bibitem{goodman} Goodman J W 1996 \textit{Introduction to Fourier Optics} (New York: McGraw-Hill Co. Inc.)

\bibitem{rayleigh} Rayleigh L 1881 On copying diffraction-gratings, and on some phenomena connected therewith \textit{Phil. Mag.} {\bf 11} 196-205

\bibitem{koza} Koza J R 1998 \textit{Genetic programming} (Cambridge: The MIT Press)

\bibitem{AG_2} Digalakisa J G and Margaritis K G 2001 On benchmarking functions for genetic algorithms \textit{International Journal of Computer Mathematics} {\bf 77} 481-506



\end{thebibliography}
\end{document}